\documentclass[aps,prl,twocolumn,a4paper,10pt,notitlepage,footinbib,nobalancelastpage,superscriptaddress,showpacs,floatfix]{revtex4-1}%
\usepackage[utf8]{inputenc}
\usepackage{graphicx,color}
\usepackage{amsmath,amssymb,latexsym}
\usepackage{mathrsfs}
\usepackage{xr}
\usepackage{lipsum}
\usepackage{verbatim}
\usepackage{mathrsfs}
\newcommand{\dmi}[1]{\textcolor{black}{#1}}%

\begin{document}

\title{Three-Dimensional Loop Extrusion}

\author{Andrea Bonato}
\address{University of Edinburgh, SUPA, School of Physics and Astronomy, Peter Guthrie Road, EH9 3FD, Edinburgh, UK }

\author{Davide Michieletto}
\thanks{davide.michieletto@ed.ac.uk}
\affiliation{University of Edinburgh, SUPA, School of Physics and Astronomy, Peter Guthrie Road, EH9 3FD, Edinburgh, UK }
\affiliation{MRC Human Genetics Unit, Institute of Genetics and Cancer, University of Edinburgh, Edinburgh EH4 2XU, UK}

\begin{abstract}
\textbf{Loop extrusion convincingly describes how certain Structural Maintenance of Chromosome (SMC) proteins mediate the formation of large DNA loops. Yet, most of the existing computational models cannot reconcile recent \emph{in vitro} observations showing that condensins can traverse each other, bypass large roadblocks and perform steps longer than its own size.  To fill this gap, we propose a three-dimensional (3D) ``trans-grabbing'' model for loop extrusion which not only reproduces the experimental features of loop extrusion by one SMC complex, but also predicts the formation of so-called ``Z-loops'' via the interaction of two or more SMCs extruding along the same DNA substrate. By performing Molecular Dynamics simulations of this model we discover that the experimentally observed asymmetry in the different types of Z-loops is a natural consequence of the DNA tethering \emph{in vitro}. Intriguingly, our model predicts this bias to disappear in absence of tethering and a third type of Z-loop, which has not yet been identified in experiments, to appear. Our model naturally explains road-block bypassing and the appearance of steps larger than the SMC size as a consequence of non-contiguous DNA grabbing. Finally, it is the first to our knowledge to address how Z-loops and bypassing might occur in a way that is broadly consistent with existing cis-only 1D loop extrusion models.
 }
\end{abstract}

\maketitle
	

\section{Introduction}

Cells exert an impressive control over genome folding to confine long chromosomes inside the small space of a nucleus. Structural maintenance of chromosome (SMC) proteins are now well known to fundamentally contribute to the large-scale folding of DNA \emph{in vivo}~\cite{Hirano2002}. Cohesin and condensin, ring-shaped SMC protein complexes, can bring together two DNA segments and form DNA loops~\cite{Nasmyth2011,Uhlmann2016,Hirano2016}. Current evidence show that yeast condensin~\cite{Ganji2018}, human cohesin~\cite{Davidson2019a,Kim2019a}, human condensin~\cite{Kong2020} and both cohesin and condensins in Xenopus egg extracts~\cite{Golfier2020}, processively extrude loops in vitro. 
At the same time, also bacterial SMCs also appear to extrude loops in vivo~\cite{Brandao2019a,Brandao2021,Anchimiuk2021}. On the other hand,, currently there is no direct evidence that yeast cohesin can extrude loops \emph{in vitro}~\cite{Stigler2016,Gutierrez-Escribano2019,Ryu2021}, although there is indirect evidence for translocation of yeast cohesin \emph{in vivo}~\cite{Glynn2004,Lengronne2004,Paldi2020}. 

The formation and growth of long DNA loops is well described by the loop extrusion model~\cite{Fudenberg2016,Goloborodko2016a,Sanborn2015a, Alipour2012, Nasmyth2011}. In this popular framework, loop extrusion factors (LEFs) such as condensin~\cite{Ganji2018} or cohesin~\cite{Davidson2019a,Kim2019a} bind DNA and stem a short loop by grabbing two contiguous DNA segments; then, they move along the DNA until they either unbind or stop (e.g., \emph{in vivo}, when they encounter a zinc-finger protein CCCTC binding factor~\cite{Phillips2009,Tang2015,Oti2016}). Evidence and models for entropic diffusion~\cite{Brackley2017prl, Davidson2016,Stigler2016,Yamamoto2017}, Brownian ratchet~\cite{Brackley2017prl,Higashi2021} or bridging~\cite{Ryu2020} have also been reported. 

Although loop extrusion convincingly describes the principles behind the formation and growth of DNA loops, the mechanical details of how SMC protein complexes extrude loops are still debated. Proposed loop extrusion models include the pumping~\cite{Marko2019,DieboldD2017}, scrunching~\cite{Terakawa2017,Takaki2020},  tethered inchworm~\cite{Nichols2018} and safety-belt~\cite{Kschonsak2017} mixed with power stroke~\cite{Nomidis2021}; all inspired by the shape and structure of the SMC complexes~\cite{RyuPre2019,DieboldD2017,Kamada2017}. 

\begin{figure*}[!]
	\begin{center}
	\includegraphics[width=0.97\textwidth]{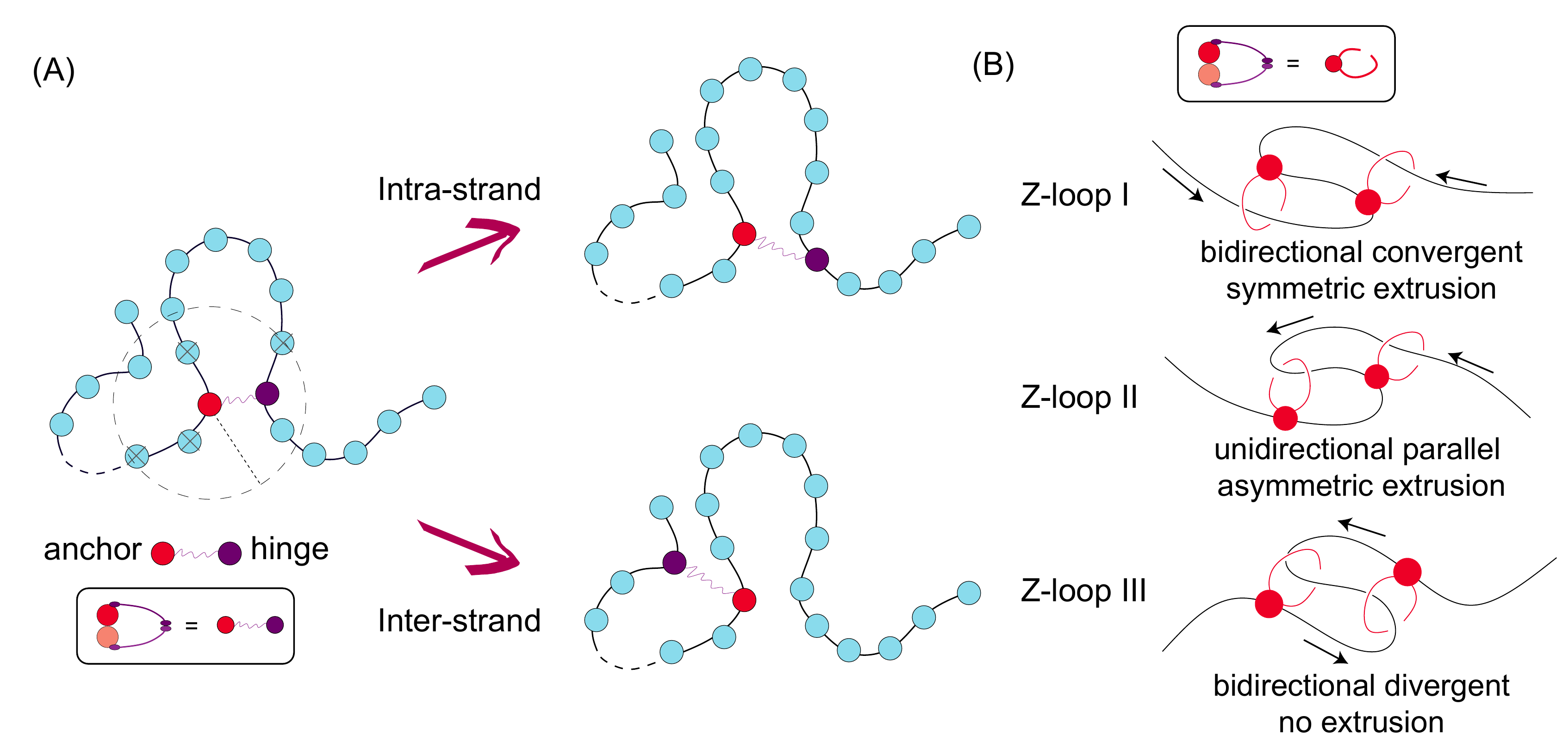}
	\end{center}
	\vspace{-0.8 cm}
	\caption{	\dmi{A three-dimensional, or ``trans-grabbing'' model for loop extrusion. (A) We model LEFs as springs connecting two polymer beads; one of these beads is denoted as the ``anchor'' that does not move, while the other as the ``hinge'' that can jump to 3D proximal but non-contiguous polymer segments. By updating the position of the hinge via a mix of intra- and inter-strand moves, the system is driven to form Z-loops. (B) We identify 3 types of Z-loops: Z-loop I ($Z_I$) has both hinges pointing outwards thus yielding symmetric extrusion;  Z-loop II ($Z_{II}$) has both hinges pointing in the same direction and thus yields asymmetric extrusion; finally, Z-loop III ($Z_{III}$) has both hinges pointing inward and displays no net extrusion. Only $Z_I$ and $Z_{II}$ were observed in experiments~\cite{Kim2020}.}
	}
	\vspace{-0.4 cm}
	\label{fig_Scheme}
\end{figure*}

Most of the previously proposed models have a common feature, they assume the extrusion to happen in cis, i.e. by reeling in contiguous DNA contour length. An exception to these is the diffusion-capture model~\cite{Cheng2015} which posits that DNA-bound SMCs can dimerize and form stable loops when they meet in trans by 3D diffusion. While both classes of models compare favorably well with experiments in vivo~\cite{Gibcus2018,Gerguri2021}, they cannot explain some recent observations in vitro. For instance, cis loop extrusion with topological entrapment cannot explain the formation of so-called ``Z-loops''~\cite{Kim2020}, i.e. non-trivial positioning of SMCs yielding cross-looping topologies (see Fig.~\ref{fig_Scheme}B), nor the bypassing of roadblocks several times the size of condensin~\cite{Pradhan2021}, nor the fact that condensin single steps can reel in DNA longer than its own size~\cite{Ryu2021stepsize}. At the same time, loop-capture mechanisms~\cite{Cheng2015,Gerguri2021} cannot explain the observed processive extrusion of condensin on tethered DNA~\cite{Ganji2018}. 

To fill this gap, and motivated by the fact that cohesin's main role is to bridge sister chromatids \emph{in vivo} (which cannot be achieved via a pure cis-looping mechanism)~\cite{Nasmyth2009,Murayama2018,Piskadlo2017a}, here we propose a trans-grabbing model for loop extrusion. By performing Molecular Dynamics simulations, we discover that Z-loops are a natural consequence of introducing occasional inter-strand capture in a standard loop extrusion model (see Fig.~\ref{fig_Scheme}). Furthermore, we find that the experimentally observed bias for Z-loop I over II (see Fig.~\ref{fig_Scheme}B) is a consequence of DNA tethering of the \emph{in vitro} assay in Ref.~\cite{Kim2020}. Finally, we find that under certain conditions, i.e. when SMCs are initiated in series and in a convergent orientation, a third type of (non-growing) Z-loop, which we dub $Z_{III}$, appears. 

We note that the novelty of our model is that it is the first to explain the formation of Z-loops in a natural way that is broadly consistent with existing models of loop extrusion. Indeed, while models accounting for non-contiguous reeling of DNA~\cite{Takaki2020,Higashi2021} or extending the original LEF model to allow bypassing have been proposed~\cite{Brandao2019a,Brandao2021}, they have not extensively explored the topological issues arising from to the interaction of multiple loop extruding factors on the same DNA substrate nor addressed the abundance and evolution of different Z-loop topologies.
 
Finally, our model also naturally rationalises recent observations of SMCs bypassing large roadblocks \emph{in vitro}~\cite{Pradhan2021} and  \emph{in vivo}~\cite{Brandao2021}, and performing steps larger than their own size~\cite{Ryu2021stepsize}. Our model -- in agreement with these recent findings -- supports the view that SMCs may perform non-topological loop extrusion composed by discrete jumps in which SMCs grab a non-contiguous DNA segment in turn reeling in the subtended contour. 

\section{Model and methods}

To simulate loop extrusion on DNA we employ a well-established coarse grained model~\cite{Fudenberg2016,Brackley2017}. We perform Molecular Dynamics simulations of a segment of torsionally relaxed DNA modelled as bead-and-spring polymer made of beads of size $\sigma=10nm \simeq 30$ base-pairs. Consecutive beads are connected by finite-extension-nonlinear-elastic (FENE) bonds, i.e.
\begin{equation}
\label{eq:Ufene}
U_{\rm FENE}(r) = \left\{
\begin{array}{lcl}
-0.5 k_F R_0^2 \ln\left(1-(r / R_0)^2\right) & \ r\le R_0 \\ \infty & \
r> R_0 &
\end{array} 
\right. \, ,
\end{equation}
where $k_F = 30k_BT/\sigma^2$ and $R_{0}=1.5\sigma$. Beads interact with each other via pure excluded volume, via a Weeks-Chandler-Andersen (WCA) potential, 
\begin{equation}\label{eq:LJ}
U_{\rm WCA}(r) = \left\{
\begin{array}{lr}
4 \epsilon \left[ \left(\frac{\sigma}{r}\right)^{12} - \left(\frac{\sigma}{r}\right)^6 + \frac14 \right] & \, r \le r_c \\
0 & \, r > r_c
\end{array} \right. \, ,
\end{equation}
where $r$ denotes the separation between the bead centers and $r_c=2^{1/6}\sigma$. The stiffness of DNA is accounted for by introducing a Kratky-Porod potential acting on triplets of consecutive beads along the polymer, 
\begin{equation}\label{Ubend}
U_{\rm B}(i,i+1,i+2) = \dfrac{k_BT l_p}{\sigma}\left[ 1 - \dfrac{\mathbf{d}_{i,i+1} \cdot \mathbf{d}_{i+1,i+2}}{d_{i,i+1}d_{i+1,i+2}} \right],
\end{equation}
where {$\mathbf{r}_i$} is the position of {$i$}-th bead,  $\mathbf{d}_{i,j}=\mathbf{r}_i-\mathbf{r}_j$ and $d_{i,j}=|\mathbf{r}_i-\mathbf{r}_j|$ are, respectively, the separation vector between beads $i$ and $j$ and its modulus. We set $l_p=5\sigma$ to achieve the known persistence length of DNA $l_p\simeq50$~nm~\cite{Bustamante1994}. We use the LAMMPS~\cite{Plimpton1995a} engine to integrate the equations of motion with implicit solvent (Langevin dynamics) with friction $\gamma=m/\tau_B$ (where the Brownian time is $\tau_B=\gamma \sigma^2/k_BT$)  is related to the thermal noise amplitude via the fluctuation-dissipation theorem. Finally, the integration step size is $10^{-4}\tau_B$.

\subsection{Implementation of the 3D loop extrusion model}
In this section we explain in detail our inter-strand loop extrusion model. In essence, we generalise the standard cis loop extrusion model~\cite{Fudenberg2016,Goloborodko2016a} by introducing the possibility of trans (3D) moves in which LEFs can grab a DNA segment that is proximal in 3D. 

Each SMC is modelled as a spring connecting two DNA beads (segments), and is described by the harmonic potential
\begin{equation}
\label{harmonic}
U_{H}=k\left(r-r_0\right)^2,
\end{equation}
where $r_0 = 1.6 \sigma$ is the resting distance between the centers of the two beads and $k = 4 k_BT/\sigma^2$ is the elastic constant. The rest length is chosen to be $r_0 = 16$ nm, comparable to the minimum size of condensin. We then fix one of the two ends of the spring (the anchor) for the duration of the simulation, whereas the other (the hinge) is periodically updated as follows.  

\begin{figure}[t!]
	\begin{center}
	\includegraphics[width=0.97\columnwidth]{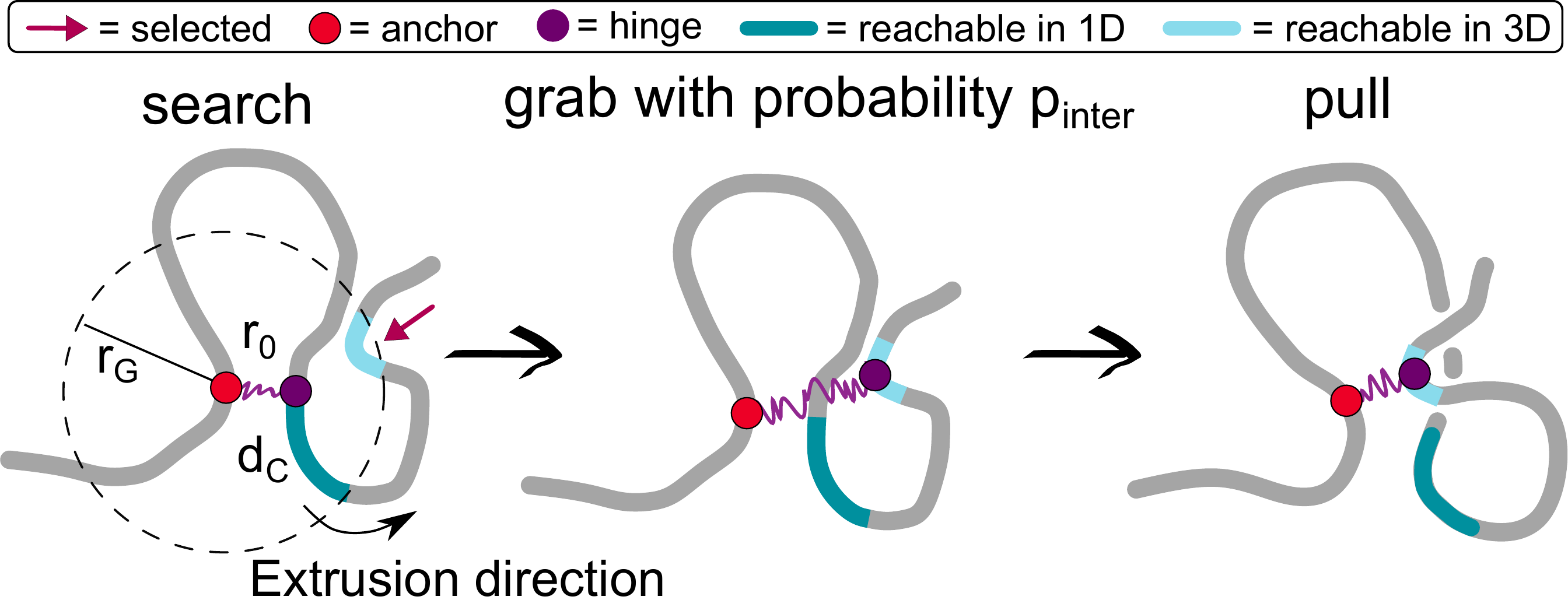}
	\end{center}
	\vspace{-0.5 cm}
	\caption{Implementation of 3D extrusion. A LEF is modelled as a spring connecting two non-contiguous beads and with rest length $r_0 = 1.6\sigma$. Every 8000 simulations steps ($\simeq 0.01$ s), we attempt to move the LEF by gathering the 3D neighbours within an Euclidean distance $r_G=3.4 \sigma$ from the anchor. Of these, the ones that fall within the $d_C=5$ nearest neighbours are classified as 1D beads. A random bead from the list of 3D ``trans'' neighbours is selected with probability $p_{inter}= 5$ $10^{-3}$ (in SI we show results with a different choice of this parameter) and a random bead from the 1D list, otherwise. Finally, we update the beads connected by the spring and evolve the equations of motion of the beads so that the spring relaxes to its equilibrium rest length $r_0$. We note that setting $p_{inter}=0$ and $d_C=2$ maps back to the standard cis-only loop extrusion model~\cite{Fudenberg2016,Sanborn2015a}.}
	\label{fig_sgp}
\end{figure}

First, after a LEF is loaded with its anchor at bead $l$ and hinge at bead $m$, we randomly choose an extrusion direction. Then, at each extrusion step, all the beads within a certain ``grabbing'' Euclidean distance $r_G=3.4\sigma=34$ nm (smaller than the size of condensin) from the anchor $l$ are identified (see Fig.~\ref{fig_sgp}). Out of those 3D proximal beads, we select two groups: the first contains all the ``cis'' beads that are within $5$ beads from the hinge $m$ and to its left or its right (according to the extrusion direction).
The second group contains the beads farther than $d_C=5$ in 1D from both the hinge $m$ and the anchor $l$. In this way we distinguish beads that are close in 3D but far in 1D and at the same time disallow back-steps which are only rarely seen in experiments. 
After the creation of these two groups, we randomly select one bead, $n$, from either the first (1D) group or, if not empty, from the second (3D) group, with a small probability $p_{inter}=5\times 10^{-3}$. Finally, we update the position of the LEF by connecting $l$ to the new selected bead $n$ (and remove the bond between $l$ and $m$). In the case that the move is in 3D, we select a new extrusion direction at random as we assume that the SMCs cannot distinguish forward/backward on a newly grabbed DNA segment. Finally, we note that setting $p_{inter}=0$ and $d_C=2$ maps our model back to standard cis-only loop extrusion models~\cite{Fudenberg2016}.

We implement this algorithm in LAMMPS~\cite{Plimpton1995a} by loading it as an external library within a C\texttt{++} program. Every 8000 integration steps (or $0.8 \tau_B$ with $\tau_B \simeq 0.011$ s), we extract the coordinates of all the beads and loop over the positions of LEFs. The bonds are then updated using the ``delete bond'' and ``create bond'' commands to connect new pair of beads as described above. Sample codes can found at \url{https://git.ecdf.ed.ac.uk/dmichiel/translefs}.

\section{Results}

\subsection{Calibration of the model using one LEF}

\begin{figure}[t]
	\begin{center}
	\includegraphics[width=0.97\columnwidth]{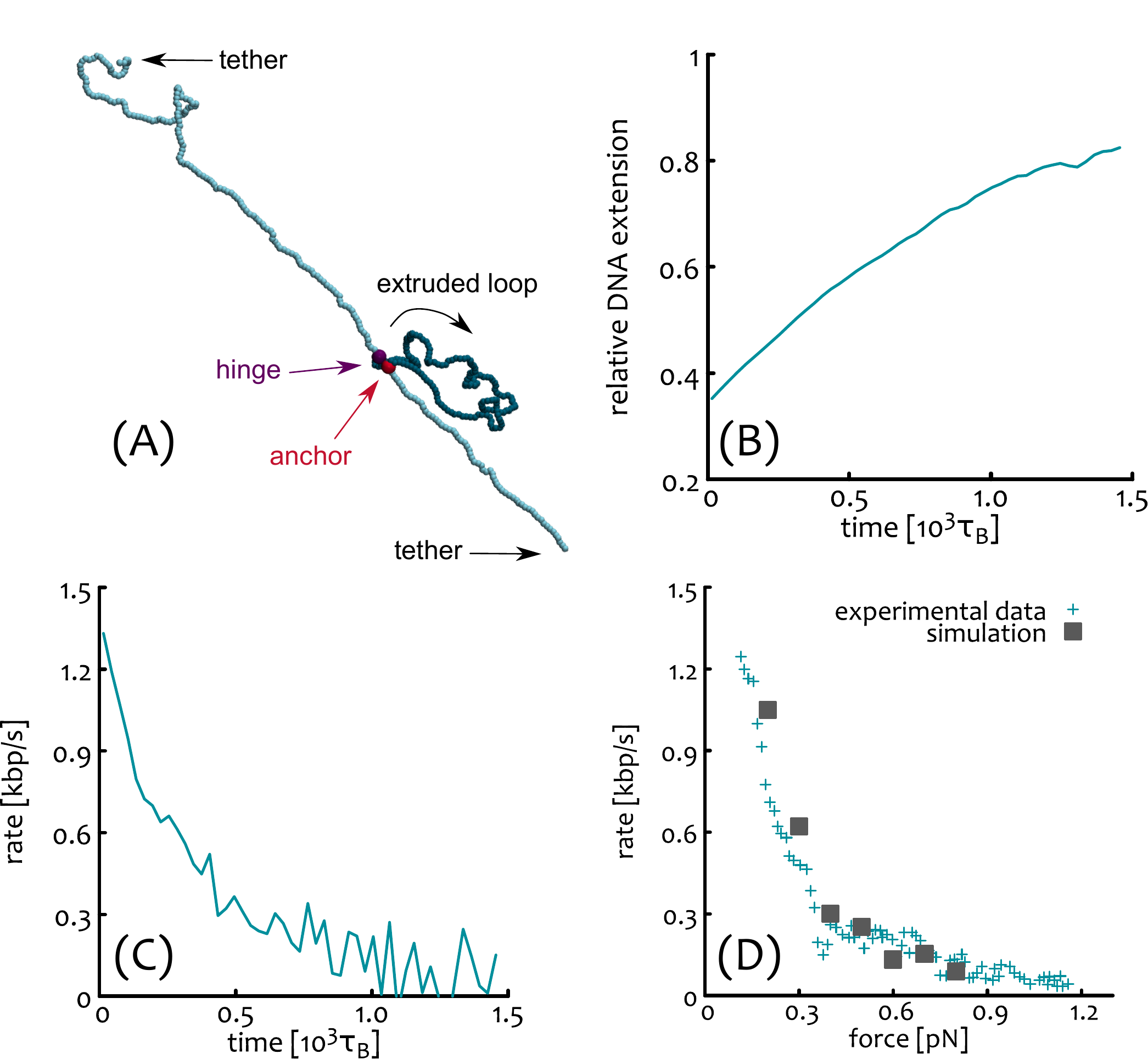}
	\end{center}
	\caption{\textbf{(A)} MD simulations of doubly tethered DNA (cyan) loaded with one asymmetric LEF (violet and red) extruding a loop (blue). \textbf{(B)} Mean relative DNA extension as a function of time. The relative extension is defined as the end-to-end distance divided by the difference between the total length of the DNA and the length of the extruded loop. \textbf{(C)} Extrusion rate as a function of time. \textbf{(D)} Extrusion rate as a function of tension: comparison between our simulations (dark grey squares) and experimental data from Ref.~\cite{Ganji2018} (cyan +).}
	\label{fig_Calibration}
\end{figure}

We first calibrate our model on doubly tethered DNA $N=400$ beads long loaded with one (asymmetric) LEF (Fig.~\ref{fig_Calibration}A). Each initial conformation is obtained by equilibrating for a long time ($10^6\tau_B$) a polymer with the constraint that the Euclidean distance between its two ends is equal to one third of its total length when pulled taut. This mimics the experimental set up in Ref.~\cite{Ganji2018} and is practically implemented by applying a force to the chain ends until they reach a distance of $N/3$ in a pre-simulation step and then by equilibrating the chain with its ends fixed in space. After equilibration, a LEF is loaded either at bead $N/6$ and with positive extrusion direction (i.e. the moving side of the spring with progressively larger bead index) or at bead $5N/6$ and with negative extrusion direction. In this way, we prevent the extrusion process to end because of the LEF reaching the end of the polymer.  

Importantly, in our simulations, the extrusion rate is not a fixed parameter, but depends on the tension experienced by the polymer during the loop extrusion as this enters in competition with the tethered ends. For instance, in Fig.~\ref{fig_Calibration}B, one can appreciate that the relative DNA extension, i.e. the ratio between the end-to-end distance  ($R_{ee}$) and the difference of DNA length ($N$) and extruded loop length ($l=l(t)$), or $R_{ee}/(N-l)$, grows in time because the difference $N-l$ becomes smaller and $R_{ee}$ is kept constant. At large times, when the loop extrusion step is balanced by the tension on the DNA, one expects to see a plateau in the relative extension. This stalling is due to the fact that in the LEF update rule, new segments are searched within a radius $r_G$ from the LEF anchor point. If the LEF bond connecting anchor and hinge is $\simeq r_G$ (notice that $r_G \simeq 2 r_0 \simeq 32$ nm), segments outside the extruded loop do not fall within the grabbing range of the LEFs. In other words, the tension along the polymer leads to stretching of the LEF bond and then to stalling via the depletion of available cis (and trans) segments that can be grabbed. 

The extrusion rate is thus computed from the simulations by taking the (discrete) time derivative of the length of the extruded loop as in experiments (Fig.~\ref{fig_Calibration}C), i.e. $rate = \partial l(t)/\partial t$.  We then plot the rate as a function of the tension by converting the extension of the polymer to force as done in Ref.~\cite{Ganji2018}, i.e. by using a tabulated conversion. 

The free parameters of the model, i.e. LEF update time $dt$, LEF spring stiffness $k$ and Brownian time $\tau_B$ were progressively tuned to closely match the experimental data in Ref.~\cite{Ganji2018}. With this calibration we thus lock in a combination of these parameters that match experimental data on single LEF extrusion on tethered DNA and leave the trans-grabbing (or inter-strand) parameter $p_{inter}$ free to be explored when studying multiple LEFs, which is the main aim of this work. Fig.~\ref{fig_Calibration}D shows the rate-force curve we ultimately obtain by simulating 100 DNA molecules with the optimal parameters (see SI for other choices of our free parameters).

\subsection{Modelling interactions of two LEFs}

\begin{figure*}[t!]	
	\includegraphics[width=.97\textwidth]{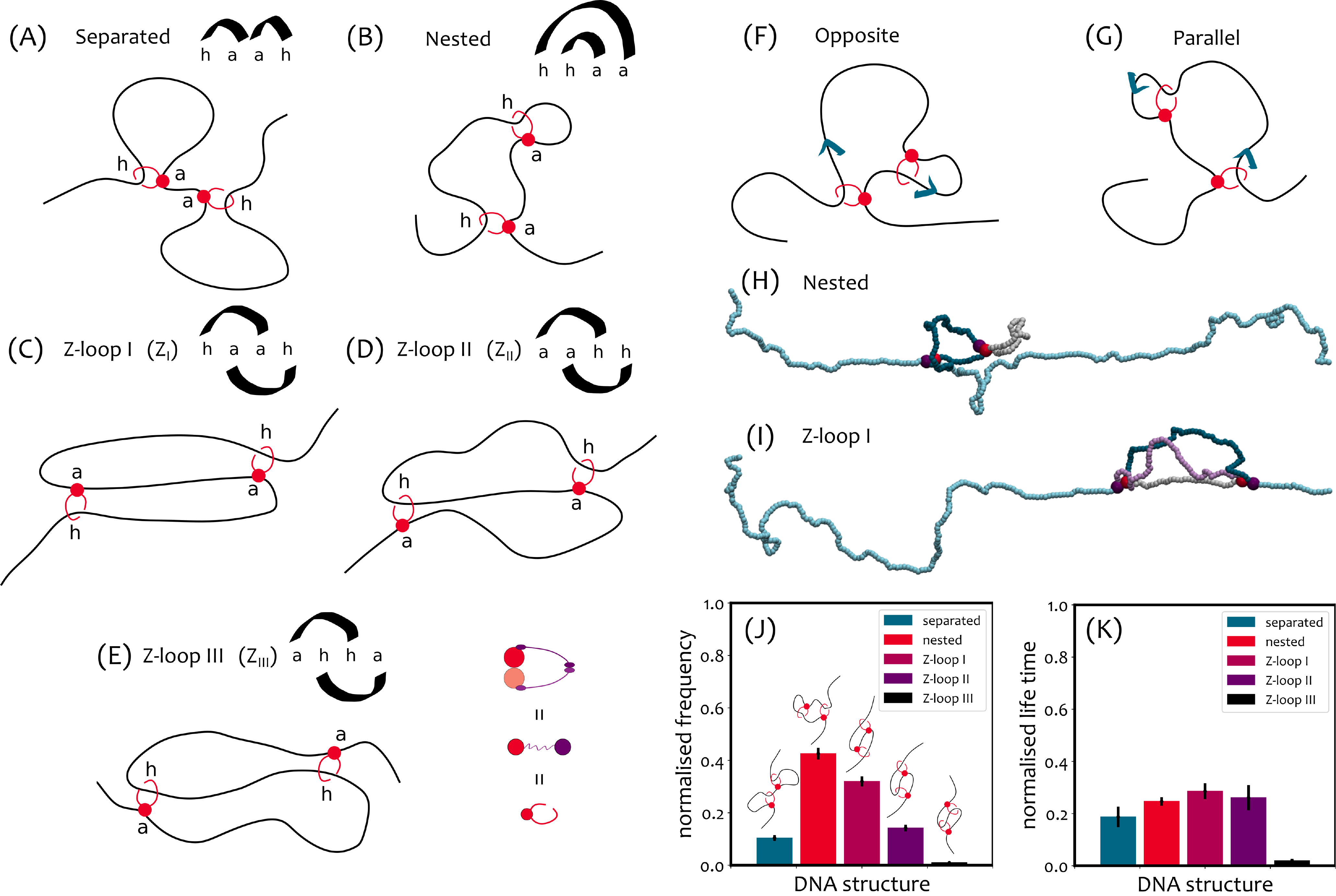}
	\caption{\textbf{(A)-(E)} A summary of the 5 Z-loop topologies observed in simulations with two LEFs (the anchored bead ``a'' and the moving hinge ``h''): \textbf{(A)} separated, \textbf{(B)} nested, \textbf{(C)} Z-loop I (this structure grows as DNA is reeled in from the outward-facing hinges); \textbf{(D)} Z-loop II (only one of the boundaries moves with respect to the structure); \textbf{(E)} Z-loop III (both boundaries of the structure are fixed as the anchors are facing inward). See also SI Fig.~S5 for a step-wise scheme on the formation of these loop topologies. \textbf{(F)(G)} Schematic illustration of the initial configuration of MD simulations with two nested LEFs. When the second LEF is loaded, its extrusion direction either opposes \textbf{(F)} or copies \textbf{(G)} the extrusion direction of the first one. Blue arrows indicate in which direction DNA is reeled inside a loop. \textbf{(H)(I)} MD simulations of doubly tethered DNA loaded with two LEFs. \textbf{(H)} When the LEFs are nested, one of the loops (grey) extruded by the two protein complexes is part of the other loop (grey+blue). \textbf{(I)} Two condensins can fold a Z-loop (type I in the figure). One of the three segments (grey) involved in a Z-loop is shared between the loops extruded by the two condensins (grey+blue and grey+lilac). \textbf{(J)(K)} Frequency \textbf{(J)} and survival times \textbf{(K)} of topological structures in simulations of two nested LEFs on doubly tethered DNA. }
	\label{fig_TwoSides_and_Structures}
\end{figure*}

After having calibrated our model using one LEF to match the experimentally observed behaviour, we now perform MD simulations of two LEFs along the same DNA substrate (see Fig.~\ref{fig_TwoSides_and_Structures}H,I). To align with the experimental set-up in Ref.~\cite{Kim2020}, we simulate doubly tethered polymers $N=400$ beads long and choose to load the LEFs in a nested state, as typically observed in experiments.  
This is done by (i) loading the first LEF immediately after equilibration in a random position and with random extrusion direction and then (ii) by attempting to load a second LEF inside the loop formed by the first some time after the simulation starts. We can distinguish between two further cases: in one half of the simulations, the two LEFs are nested and extruding in the same direction (Fig.~\ref{fig_TwoSides_and_Structures}F) while in the second half the LEFs have opposite extrusion direction (Fig.~\ref{fig_TwoSides_and_Structures}G).

To check the evolution of the polymer topology over time, we reconstruct the position of the ends of the LEFs and draw a corresponding arch diagram for each observed configuration. As shown in Fig.~\ref{fig_TwoSides_and_Structures}A-E, we observe separated, nested and 3 types of Z-loop topologies. It should be noted that all these structures are non-equilibrium topologies since the LEFs are unidirectionally moving along the DNA and thus displaying an absorbing non-extruding state at large times. We take our simulations total runtime to be typically shorter than the time it takes for the LEFs to reach the absorbing state. Additionally, the same topology can appear more than once for each simulation: it can form, undo and eventually form again at a later time. A step-wise scheme of loop formation and disassembly is reported in SI, Fig.~S5. We stress that it is not the loops themselves that are undone (our LEFs are never unloaded from the substrate), but the Zloop topologies formed by the interaction of the two loops that evolve in time.

To best compare with experiments, we align to the method of Ref.~\cite{Kim2020} and record (i) the number of times we observe a given loop topology and (ii) its relative survival time over a fixed total simulation runtime. Fixing a total runtime is important as the final states are absorbing and would therefore dominate the spectrum of topologies in the very large time limit.

In Fig.~\ref{fig_TwoSides_and_Structures}J,K we show the relative frequency and survival times of the 5 different topologies. 
As one can notice, while the nested state is the most likely topology, Z-loops I and II are also significant. Remarkably, we observe a spontaneous asymmetry in topologies, whereby Z-loop I ($Z_I$) is more than twice as likely to form than Z-loop II ($Z_{II}$), i.e. $Z_I/Z_{II}\simeq 2.5$. This asymmetry emerges naturally, without imposing any bias favouring the formation of a particular Z-loop type.
Additionally, it is in good agreement with experiments, as they report $Z_I/Z_{II} \simeq 3$~\cite{Kim2020}. We find that this asymmetry is assay and tension dependent, and that singly-tethered DNA display a far weaker bias in $Z_I/Z_{II}$ (see next section). 

We finally highlight that while we observe Z-loop III ($Z_{III}$) in simulations (Fig.~\ref{fig_TwoSides_and_Structures}E), they have not been detected in experiments~\cite{Kim2020}. Accordingly, the frequency with which they appear in our simulations is tens of times smaller than the other Z-loop topologies. Again, we predict that this result is assay-dependent (see below).

\begin{figure}[t!]	
\includegraphics[width=.97\columnwidth]{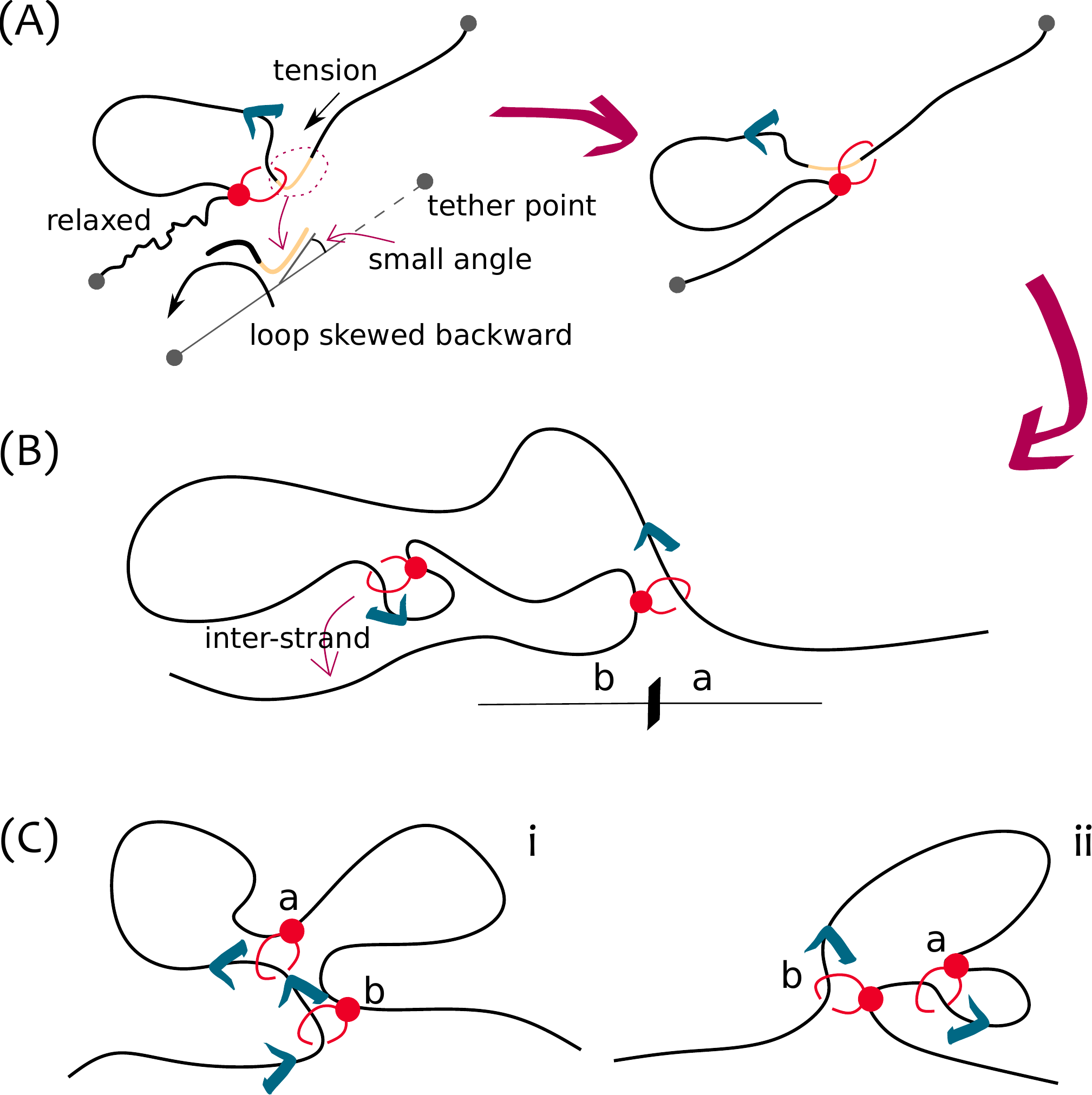}
\caption{Asymmetry in the frequency of $Z_I$ and $Z_{II}$ on DNA with two tethered ends. \textbf{(A)} At each extrusion step, the segment of DNA (cream) on the intake side is likely to be pulled taut and roughly parallel to the line connecting the two ends of the tethered DNA. The anchor-side part is instead relaxed as no tension is applied. \textbf{(B)} This effective alignment biases the extruded loop to fold over the anchor-side of the most external LEF. In turn, this favours the nested LEF to grab segments of DNA from over the anchor side (b) rather than over the hinge side (a)  (see Fig.~\ref{fig_Calibration}A). \textbf{(C) \textbf{(i)}} If the two nested LEFs are extruding in the same direction and the internal LEF, labelled a, approaches the other one (b), a $Z_{II}$ is formed when a's hinge jumps other b's. When the jump is attempted, however, b's hinge is running away from a's. \textbf{(C) (ii)} If instead the LEFs are extruding in opposite directions, a $Z_I$ is formed when a's hinge jumps other b's anchor, which does not move. Move \textbf{(ii)} (yielding $Z_{I}$) is more likely to succeed due to the slower dynamics of the polymer near a LEF anchor.}
\label{fig_Asymmetry}
\end{figure}

\subsection{Z-loop asymmetry is due to tension and assay geometry}

We can think of at least two reasons that may bias the formation of $Z_I$ over $Z_{II}$. As sketched in Fig.~\ref{fig_Asymmetry}, because of the geometry of the doubly-tethered assay, more frequent inter-grabbing events are expected to occur on the side of the anchor of the outermost LEF (labelled b in Fig.~\ref{fig_Asymmetry}A,B). This is because the extrusion is unidirectional and the DNA, being tethered, has a preferred structural direction in the 3D space (the line passing through the two fixed ends). Each time a loop extruded by the external LEF grows, a new segment of DNA is brought inside the loop. Since the DNA is tethered, and especially when the relative DNA extension is close to one, the angle between this newly added segment and the geometric line passing through the two ends of the DNA is small (see Fig.~\ref{fig_Asymmetry}A).

As shown in Fig.~\ref{fig_Asymmetry}A (see also snapshots in Figs.~\ref{fig_Calibration}A and \ref{fig_TwoSides_and_Structures}H), as the extrusion goes on, the geometry of the extruding complexes biases the extruded DNA loop to fold over the anchor (where there is no force applied by the LEF). As a consequence, due to closer proximity, it then becomes easier for the nested LEF to grab a segment behind the anchor of the external LEF (side b) in turn forming a $Z_I$ loop (Fig.~\ref{fig_Asymmetry}B), rather than grabbing a segment over the hinge of the other LEF (side a) to eventually form a $Z_{II}$ loop.

Moreover, when two nested LEFs are both extruding in the same direction along the DNA, one of them is ``running away'' from the other. As shown in Fig.~\ref{fig_Asymmetry}C, this hinders the hinge of the nested LEF to jump over the other when folding a $Z_{II}$. This suggests that directionality of LEFs also play a role in the balance of Z-loop topologies (see also SI).

\begin{figure}[t!]	
	\includegraphics[width=.97\columnwidth]{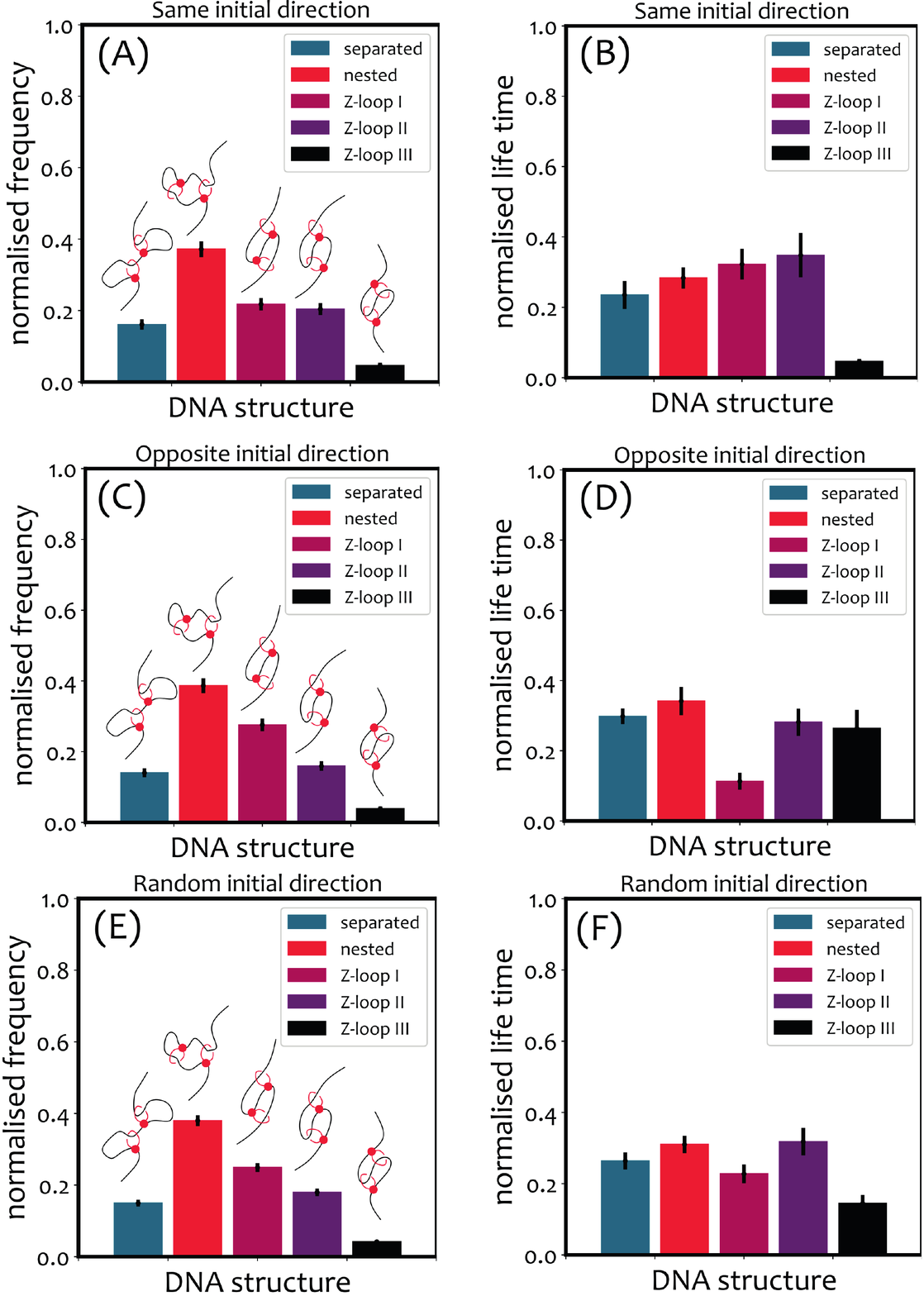}
	\caption{Results of simulations of nested LEFs on singly-tethered DNA. \textbf{(A)(C)(E)} Frequency of topologies folded by two LEFs loaded with the \textbf{(A)} same or \textbf{(C)} opposite initial directions. \textbf{(B)(D)(F)} Mean normalised life time of the loop topologies.}
	\label{fig_OneSide}
\end{figure}

\subsection{Singly- vs doubly-tethered DNA}

We hypothesise that the tethering of the DNA may bias the local conformation of extruded loops especially in the limit where the relative extension is close to one. To test this, we simulate two nested LEFs on singly-tethered DNA. 

While this setup was considered in experiments~\cite{Kim2020}, the imaging resolution did not allow a clear interpretation of the looped topologies.  On the contrary, in our simulations we can always precisely classify and distinguish different Z-loop topologies. As before, we can also distinguish between nested LEFs extruding in parallel versus opposite directions. In Fig.~\ref{fig_OneSide} we show that, compared with the doubly-tethered case, the frequencies at which $Z_I$ and $Z_{II}$ are closer to each other. We argue that this should also hold for free DNA. The fraction of $Z_{III}$ is now larger than before, but still significantly lower than $Z_{I,II}$; yet, the corresponding mean (normalised) life time is substantially larger (see Fig.~\ref{fig_OneSide}D), meaning that it forms relatively rarely, but when it does, it is a stable topology.

We note that $Z_{III}$ loops are clearly stable due to their topology. The anchors point outward therefore making it a non-extruding Z-loop structure. Without polymer fluctuations, a $Z_{III}$ should be an absorbing state for the system as it cannot evolve by pure 1D extrusion. Thanks to polymer fluctuations and 3D moves this topology can come undone by, e.g., bypassing an hinge over the other anchor (thus going into a nested state). 

These results confirm that, as hypothesised in the previous section, tethering the ends of DNA favours the formation of $Z_I$. We may thus conjecture that the statistics of Z-loop formation \emph{in vivo} could differ from those observed \emph{in vitro}~\cite{Kim2020}. To test our prediction, that DNA tension and assay affect the statistics of Z-loops, more \emph{in vitro} experiments with a different choice of set ups are needed. 

\subsection{Z-loop formation depends on initial loading}

\begin{figure}[t!]	
	\includegraphics[width=.95\columnwidth]{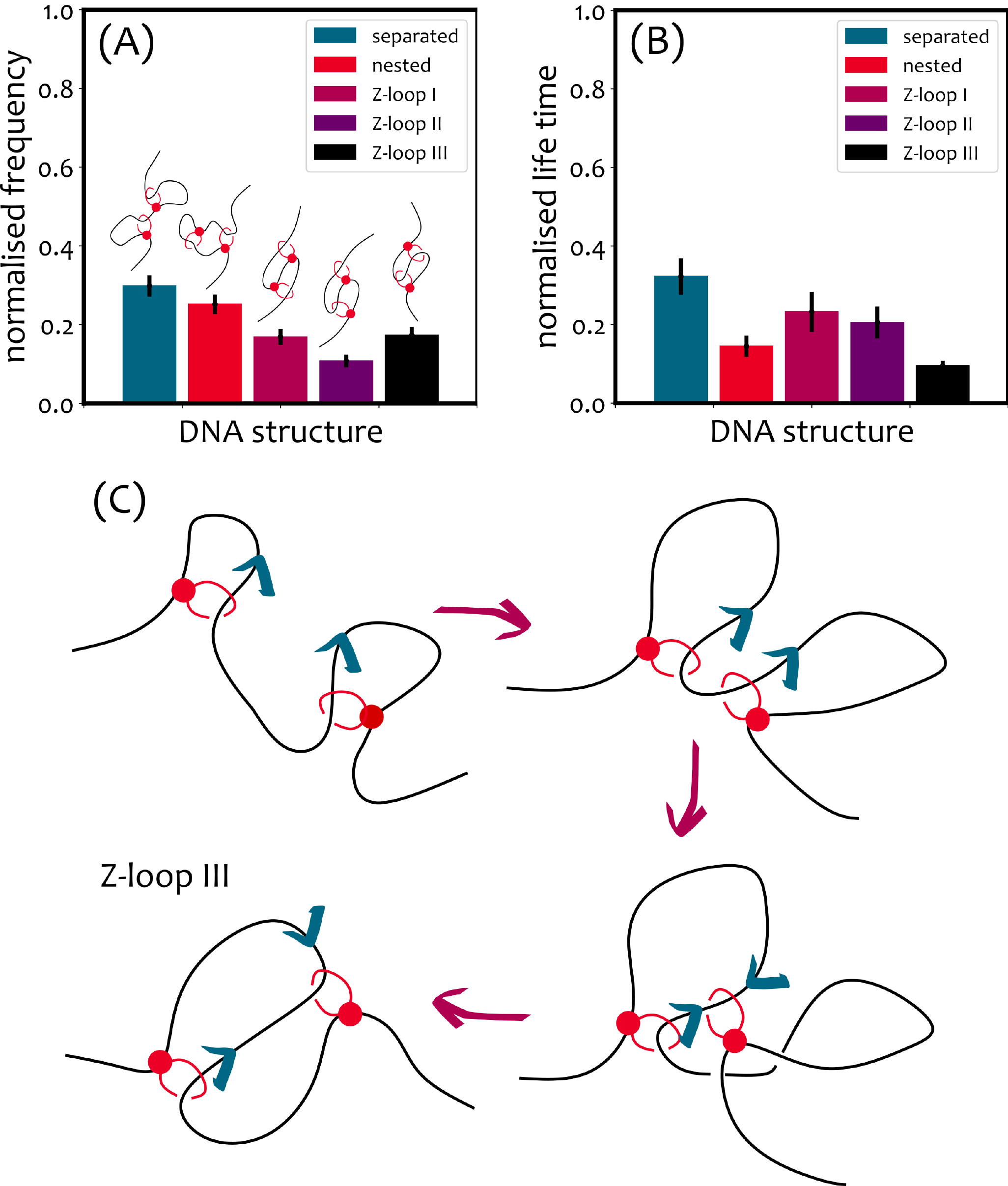}
	\caption{\textbf{(A)(B)} Results of simulations of two initially separated LEFs loaded on doubly-tethered DNA with opposite initial extrusion directions. Frequency \textbf{(A)} and mean normalised life time \textbf{(B)} of the topologies. \textbf{(C)} Scheme of the formation of a $Z_{III}$ in simulations with two initially separated LEFs and opposite extrusion directions. $Z_{III}$ is formed if one of the hinges jumps over the other.}
	\label{fig_Series}
\end{figure}

Until now we have only considered situations in which the LEFs started from a nested configuration. This is frequently observed in experiments, as condensins favour binding to bent or supercoiled substrates~\cite{Kimura1997,Kimura1999,Kim2021}. On the other hand, in our simulations we can also study the case in which LEFs start in a serial (separated) state rather than nested.  

To this end we load two LEFs at random on an equilibrated doubly-tethered polymer and impose them to have opposite extrusion directions. We simulate 50 polymers and collect the results in Fig.~\ref{fig_Series}: the measured frequencies (Fig.~\ref{fig_Series}A) and mean normalised life times (Fig.~\ref{fig_Series}B) differ substantially from those estimated for two initially nested LEFs. Specifically, the frequency of $Z_I$ remains larger than $Z_{II}$, however the fraction of $Z_{III}$ is much more significant, as it rivals the fraction of $Z_{I}$. This is not surprising: the formation of $Z_{III}$ is expected when the hinges of two LEFs meet and one jumps over the other (see Fig.~\ref{fig_Series}C).
 
\section{Conclusions}

Motivated by the observation that SMCs can tie Z-loops, bypass roadblocks and take steps larger than their own size, we have proposed a generalisation of the original cis-only loop extrusion picture to allow ``trans'' grabbing of 3D proximal DNA segments.

First, we have calibrated our model to match the experimental force-rate curve of  a single condensin extruding loops on doubly-tethered $\lambda$-DNA~\cite{Ganji2018}. Then, using the so-determined parameters, we have studied the balance and stability of loop topologies that can emerge when two LEFs are extruding on the same DNA substrate.  
 
Importantly, we have identified 5 different loop topologies: separated (or serial), nested (or parallel) and 3 types of Z-loops. These are most clearly classified in terms of their net extrusion capability: $Z_I$ can extrude from both boundaries, $Z_{II}$ can extrude from only one of the two boundaries while $Z_{III}$ (never seen in experiments so far) cannot extrude loops as the anchors of the LEFs are facing outward. We highlight that the ratio of $Z_{I}/Z_{II}$ observed in our simulations is in good agreement with the one in recent experiments~\cite{Kim2020} ($Z_{I}/Z_{II} \simeq 2.5-3$). More interestingly, this asymmetry emerges spontaneously  and is not introduced by hand in our simulation. We have shown that this asymmetry is assay-dependent and that loop extrusion on doubly-tethered DNA induces geometric conformations that favours geometries that favour the formation of $Z_I$ loops (Fig.~\ref{fig_Asymmetry}B). To test this hypothesis, we studied the formation of Z-loops on singly-tethered DNA and indeed observed that $Z_I$ are in this case as likely as $Z_{II}$. The singly-tethered DNA assay was also performed in Ref.~\cite{Kim2020} but due to the limits of optical resolution the authors could not classify the type of the Z-loops formed. 

We have also discovered a new type of Z-loops ($Z_{III}$) which is expected to be rare, but very long lived, loop topology due to the fact that the anchors point outward thereby rendering it a non-extruding form of Z-loop. We find that it appears frequently in the case that LEFs are initially separated and extruding in opposite directions. This is a prediction of our simulations that could be tested in experiments by forcing the loading of condensins on two separated sites rather than nested.  

We conclude highlighting that the precise mechanics of DNA loop extrusion is still highly debated. Here we have not attempted to give a precise biochemical characterisation of the steps involved with extrusion (for this see e.g. Refs.~\cite{Marko2019,Nomidis2019}). Instead, we have focused on relaxing the assumption of cis extrusion, which is now difficult to reconcile with recent experimental findings (such as condesin-condensin and condensin-roadblock bypassing). Additionally, Ryu et al~\cite{Ryu2021stepsize} using high-resolution magnetic tweezers recently observed that condensin collapses DNA in discrete steps, and that it can reel in more than 100 base pairs in a single step. These observations may not be compatible with cis-only loop extrusion and it is clear that new models models of loop extrusion are needed.

We argue that two types of experiments may help to disentangle this problem. First, single molecule set ups using two entangled DNA molecules in an ``X'' configuration as in Ref.~\cite{Gutierrez-Escribano2019} could provide clear evidence of SMCs jumping/bridging in trans while extruding. Our model would predict that the complex should try, if close enough, to grab a sister DNA strand while being anchored to the first. Second, experiments performed on bulk solutions of entangled DNA with and without SMCs may display different viscoelastic properties depending on whether SMCs can extrude only in cis (1D) or also in trans (3D). We hope to report on both these types of assays in the near future. 

In conclusion, the novelty of our work is that, to the best of our knowledge, it is the first to rationalise recent observations (Z-loops, roadblock bypassing and large condesin step sizes) in a manner that is compatible with existing successful models of loop extrusion. In fact, our model is a generalisation of cis loop extrusion that allows occasional 3D grabbing events. Additionally, and perhaps most importantly, we feel that our work suggests that more comprehensive models of loop extrusion may be needed to explain the organisation of DNA \emph{in vivo}. Currently, many models for interphase and mitotic genome organisation employ cis loop extrusion~\cite{Fudenberg2016,Goloborodko2016a}, yet it is well known that cohesin bridges sister chromatids together and without it the mitotic structure would fall apart~\cite{Nasmyth2009,Murayama2018,Piskadlo2017a}. We thus argue that some SMCs, and in some conditions, may be able to activate a trans-loop extrusion mechanisms, or bridging mode~\cite{Ryu2021}, with profound consequences on the organisation and dynamics of genomes \emph{in vivo}.

\section{Acknowledgements}
DM is a Royal Society University Research Fellow. This work was supported by the ERC (TAP, 947918). The code to simulate inter-strand loop extrusion was developed by AB and is deposited open access at \url{https://git.ecdf.ed.ac.uk/dmichiel/translefs}. 

\section{Author Contribution}
DM and AB designed research. AB developed the code, performed the simulations and analysis. DM and AB wrote the paper.

\section{Competing Interests}
The authors declare no competing interests

\bibliography{library}

\providecommand*{\mcitethebibliography}{\thebibliography}
\csname @ifundefined\endcsname{endmcitethebibliography}
{\let\endmcitethebibliography\endthebibliography}{}
\begin{mcitethebibliography}{55}
\providecommand*{\natexlab}[1]{#1}
\providecommand*{\mciteSetBstSublistMode}[1]{}
\providecommand*{\mciteSetBstMaxWidthForm}[2]{}
\providecommand*{\mciteBstWouldAddEndPuncttrue}
  {\def\EndOfBibitem{\unskip.}}
\providecommand*{\mciteBstWouldAddEndPunctfalse}
  {\let\EndOfBibitem\relax}
\providecommand*{\mciteSetBstMidEndSepPunct}[3]{}
\providecommand*{\mciteSetBstSublistLabelBeginEnd}[3]{}
\providecommand*{\EndOfBibitem}{}
\mciteSetBstSublistMode{f}
\mciteSetBstMaxWidthForm{subitem}
{(\emph{\alph{mcitesubitemcount}})}
\mciteSetBstSublistLabelBeginEnd{\mcitemaxwidthsubitemform\space}
{\relax}{\relax}

\bibitem[Hirano and Hirano(2002)]{Hirano2002}
M.~Hirano and T.~Hirano, \emph{EMBO J.}, 2002, \textbf{21}, 5733--5744\relax
\mciteBstWouldAddEndPuncttrue
\mciteSetBstMidEndSepPunct{\mcitedefaultmidpunct}
{\mcitedefaultendpunct}{\mcitedefaultseppunct}\relax
\EndOfBibitem
\bibitem[Nasmyth(2011)]{Nasmyth2011}
K.~Nasmyth, \emph{Nat. Cell Biol.}, 2011, \textbf{13}, 1170--1177\relax
\mciteBstWouldAddEndPuncttrue
\mciteSetBstMidEndSepPunct{\mcitedefaultmidpunct}
{\mcitedefaultendpunct}{\mcitedefaultseppunct}\relax
\EndOfBibitem
\bibitem[Uhlmann(2016)]{Uhlmann2016}
F.~Uhlmann, \emph{Nat. Rev. Mol. Cell. Biol.}, 2016, \textbf{17},
  399--412\relax
\mciteBstWouldAddEndPuncttrue
\mciteSetBstMidEndSepPunct{\mcitedefaultmidpunct}
{\mcitedefaultendpunct}{\mcitedefaultseppunct}\relax
\EndOfBibitem
\bibitem[Hirano(2016)]{Hirano2016}
T.~Hirano, \emph{Cell}, 2016, \textbf{164}, 847--857\relax
\mciteBstWouldAddEndPuncttrue
\mciteSetBstMidEndSepPunct{\mcitedefaultmidpunct}
{\mcitedefaultendpunct}{\mcitedefaultseppunct}\relax
\EndOfBibitem
\bibitem[Ganji \emph{et~al.}(2018)Ganji, Shaltiel, Bisht, Kim, Kalichava,
  Haering, and Dekker]{Ganji2018}
M.~Ganji, I.~A. Shaltiel, S.~Bisht, E.~Kim, A.~Kalichava, C.~H. Haering and
  C.~Dekker, \emph{Science}, 2018, \textbf{360}, 102--105\relax
\mciteBstWouldAddEndPuncttrue
\mciteSetBstMidEndSepPunct{\mcitedefaultmidpunct}
{\mcitedefaultendpunct}{\mcitedefaultseppunct}\relax
\EndOfBibitem
\bibitem[Davidson \emph{et~al.}(2019)Davidson, Bauer, Goetz, Tang, Wutz, and
  Peters]{Davidson2019a}
I.~F. Davidson, B.~Bauer, D.~Goetz, W.~Tang, G.~Wutz and J.-M. Peters,
  \emph{Science}, 2019, \textbf{366}, 1338--1345\relax
\mciteBstWouldAddEndPuncttrue
\mciteSetBstMidEndSepPunct{\mcitedefaultmidpunct}
{\mcitedefaultendpunct}{\mcitedefaultseppunct}\relax
\EndOfBibitem
\bibitem[Kim \emph{et~al.}(2019)Kim, Shi, Zhang, Finkelstein, and Yu]{Kim2019a}
Y.~Kim, Z.~Shi, H.~Zhang, I.~J. Finkelstein and H.~Yu, \emph{Science}, 2019,
  \textbf{366}, 1345--1349\relax
\mciteBstWouldAddEndPuncttrue
\mciteSetBstMidEndSepPunct{\mcitedefaultmidpunct}
{\mcitedefaultendpunct}{\mcitedefaultseppunct}\relax
\EndOfBibitem
\bibitem[Kong \emph{et~al.}(2020)Kong, Cutts, Pan, Beuron, Kaliyappan, Xue,
  Morris, Musacchio, Vannini, and Greene]{Kong2020}
M.~Kong, E.~E. Cutts, D.~Pan, F.~Beuron, T.~Kaliyappan, C.~Xue, E.~P. Morris,
  A.~Musacchio, A.~Vannini and E.~C. Greene, \emph{Molecular Cell}, 2020,
  \textbf{79}, 99--114.e9\relax
\mciteBstWouldAddEndPuncttrue
\mciteSetBstMidEndSepPunct{\mcitedefaultmidpunct}
{\mcitedefaultendpunct}{\mcitedefaultseppunct}\relax
\EndOfBibitem
\bibitem[Golfier \emph{et~al.}(2020)Golfier, Quail, Kimura, and
  Brugu{\'{e}}s]{Golfier2020}
S.~Golfier, T.~Quail, H.~Kimura and J.~Brugu{\'{e}}s, \emph{eLife}, 2020,
  \textbf{9}, 1--34\relax
\mciteBstWouldAddEndPuncttrue
\mciteSetBstMidEndSepPunct{\mcitedefaultmidpunct}
{\mcitedefaultendpunct}{\mcitedefaultseppunct}\relax
\EndOfBibitem
\bibitem[Brand{\~{a}}o \emph{et~al.}(2019)Brand{\~{a}}o, Paul, van~den Berg,
  Rudner, Wang, and Mirny]{Brandao2019a}
H.~B. Brand{\~{a}}o, P.~Paul, A.~A. van~den Berg, D.~Z. Rudner, X.~Wang and
  L.~A. Mirny, \emph{Proceedings of the National Academy of Sciences of the
  United States of America}, 2019, \textbf{116}, 20489--20499\relax
\mciteBstWouldAddEndPuncttrue
\mciteSetBstMidEndSepPunct{\mcitedefaultmidpunct}
{\mcitedefaultendpunct}{\mcitedefaultseppunct}\relax
\EndOfBibitem
\bibitem[Brand{\~{a}}o \emph{et~al.}(2021)Brand{\~{a}}o, Ren, Karaboja, Mirny,
  and Wang]{Brandao2021}
H.~B. Brand{\~{a}}o, Z.~Ren, X.~Karaboja, L.~A. Mirny and X.~Wang, \emph{Nature
  Structural and Molecular Biology}, 2021, \textbf{28}, 642--651\relax
\mciteBstWouldAddEndPuncttrue
\mciteSetBstMidEndSepPunct{\mcitedefaultmidpunct}
{\mcitedefaultendpunct}{\mcitedefaultseppunct}\relax
\EndOfBibitem
\bibitem[Anchimiuk \emph{et~al.}(2021)Anchimiuk, Lioy, Bock, Minnen, Boccard,
  and Gruber]{Anchimiuk2021}
A.~Anchimiuk, V.~S. Lioy, F.~P. Bock, A.~Minnen, F.~Boccard and S.~Gruber,
  \emph{eLife}, 2021, \textbf{10}, 1--22\relax
\mciteBstWouldAddEndPuncttrue
\mciteSetBstMidEndSepPunct{\mcitedefaultmidpunct}
{\mcitedefaultendpunct}{\mcitedefaultseppunct}\relax
\EndOfBibitem
\bibitem[Stigler \emph{et~al.}(2016)Stigler, {\c{C}}amdere, Koshland, and
  Greene]{Stigler2016}
J.~Stigler, G.~{\"{O}}. {\c{C}}amdere, D.~E. Koshland and E.~C. Greene,
  \emph{Cell Rep.}, 2016, \textbf{15}, 988--998\relax
\mciteBstWouldAddEndPuncttrue
\mciteSetBstMidEndSepPunct{\mcitedefaultmidpunct}
{\mcitedefaultendpunct}{\mcitedefaultseppunct}\relax
\EndOfBibitem
\bibitem[Gutierrez-Escribano \emph{et~al.}(2019)Gutierrez-Escribano, Newton,
  Llaur{\'{o}}, Huber, Tanasie, Davy, Aly, Aramayo, Montoya, Kramer, Stigler,
  Rueda, and Aragon]{Gutierrez-Escribano2019}
P.~Gutierrez-Escribano, M.~D. Newton, A.~Llaur{\'{o}}, J.~Huber, L.~Tanasie,
  J.~Davy, I.~Aly, R.~Aramayo, A.~Montoya, H.~Kramer, J.~Stigler, D.~S. Rueda
  and L.~Aragon, \emph{Science Advances}, 2019, \textbf{5}, 1--16\relax
\mciteBstWouldAddEndPuncttrue
\mciteSetBstMidEndSepPunct{\mcitedefaultmidpunct}
{\mcitedefaultendpunct}{\mcitedefaultseppunct}\relax
\EndOfBibitem
\bibitem[Ryu \emph{et~al.}(2021)Ryu, Bouchoux, Liu, Kim, Minamino, de~Groot,
  Katan, Bonato, Marenduzzo, Michieletto, Uhlmann, and Dekker]{Ryu2021}
J.-K. Ryu, C.~Bouchoux, H.~W. Liu, E.~Kim, M.~Minamino, R.~de~Groot, A.~J.
  Katan, A.~Bonato, D.~Marenduzzo, D.~Michieletto, F.~Uhlmann and C.~Dekker,
  \emph{Science Advances}, 2021, \textbf{7}, eabe5905\relax
\mciteBstWouldAddEndPuncttrue
\mciteSetBstMidEndSepPunct{\mcitedefaultmidpunct}
{\mcitedefaultendpunct}{\mcitedefaultseppunct}\relax
\EndOfBibitem
\bibitem[Glynn \emph{et~al.}(2004)Glynn, Megee, Yu, Mistrot, Unal, Koshland,
  DeRisi, and Gerton]{Glynn2004}
E.~F. Glynn, P.~C. Megee, H.~G. Yu, C.~Mistrot, E.~Unal, D.~E. Koshland, J.~L.
  DeRisi and J.~L. Gerton, \emph{PLoS Biology}, 2004, \textbf{2}, year\relax
\mciteBstWouldAddEndPuncttrue
\mciteSetBstMidEndSepPunct{\mcitedefaultmidpunct}
{\mcitedefaultendpunct}{\mcitedefaultseppunct}\relax
\EndOfBibitem
\bibitem[Lengronne \emph{et~al.}(2004)Lengronne, Katou, Mori, Yokabayashi,
  Kelly, Ito, Watanabe, Shirahige, and Uhlmann]{Lengronne2004}
A.~Lengronne, Y.~Katou, S.~Mori, S.~Yokabayashi, G.~P. Kelly, T.~Ito,
  Y.~Watanabe, K.~Shirahige and F.~Uhlmann, \emph{Nature}, 2004, \textbf{430},
  573--578\relax
\mciteBstWouldAddEndPuncttrue
\mciteSetBstMidEndSepPunct{\mcitedefaultmidpunct}
{\mcitedefaultendpunct}{\mcitedefaultseppunct}\relax
\EndOfBibitem
\bibitem[Paldi \emph{et~al.}(2020)Paldi, Alver, Robertson, Schalbetter, Kerr,
  Kelly, Baxter, Neale, and Marston]{Paldi2020}
F.~Paldi, B.~Alver, D.~Robertson, S.~A. Schalbetter, A.~Kerr, D.~A. Kelly,
  J.~Baxter, M.~J. Neale and A.~L. Marston, \emph{Nature}, 2020, \textbf{582},
  119--123\relax
\mciteBstWouldAddEndPuncttrue
\mciteSetBstMidEndSepPunct{\mcitedefaultmidpunct}
{\mcitedefaultendpunct}{\mcitedefaultseppunct}\relax
\EndOfBibitem
\bibitem[Fudenberg \emph{et~al.}(2016)Fudenberg, Imakaev, Lu, Goloborodko,
  Abdennur, and Mirny]{Fudenberg2016}
G.~Fudenberg, M.~Imakaev, C.~Lu, A.~Goloborodko, N.~Abdennur and L.~A. Mirny,
  \emph{Cell Rep.}, 2016, \textbf{15}, 2038--2049\relax
\mciteBstWouldAddEndPuncttrue
\mciteSetBstMidEndSepPunct{\mcitedefaultmidpunct}
{\mcitedefaultendpunct}{\mcitedefaultseppunct}\relax
\EndOfBibitem
\bibitem[Goloborodko \emph{et~al.}(2016)Goloborodko, Marko, and
  Mirny]{Goloborodko2016a}
A.~Goloborodko, J.~F. Marko and L.~A. Mirny, \emph{Biophys. J.}, 2016,
  \textbf{110}, 2162--2168\relax
\mciteBstWouldAddEndPuncttrue
\mciteSetBstMidEndSepPunct{\mcitedefaultmidpunct}
{\mcitedefaultendpunct}{\mcitedefaultseppunct}\relax
\EndOfBibitem
\bibitem[Sanborn \emph{et~al.}(2015)Sanborn, Rao, Huang, Durand, Huntley,
  Jewett, Bochkov, Chinnappan, Cutkosky, Li, Geeting, Gnirke, Melnikov,
  McKenna, Stamenova, Lander, and Aiden]{Sanborn2015a}
A.~L. Sanborn, S.~S.~P. Rao, S.-C. Huang, N.~C. Durand, M.~H. Huntley, A.~I.
  Jewett, I.~D. Bochkov, D.~Chinnappan, A.~Cutkosky, J.~Li, K.~P. Geeting,
  A.~Gnirke, A.~Melnikov, D.~McKenna, E.~K. Stamenova, E.~S. Lander and E.~L.
  Aiden, \emph{Proc. Natl. Acad. Sci. USA}, 2015, \textbf{112}, 201518552\relax
\mciteBstWouldAddEndPuncttrue
\mciteSetBstMidEndSepPunct{\mcitedefaultmidpunct}
{\mcitedefaultendpunct}{\mcitedefaultseppunct}\relax
\EndOfBibitem
\bibitem[Alipour and Marko(2012)]{Alipour2012}
E.~Alipour and J.~F. Marko, \emph{Nucleic Acids Res.}, 2012, \textbf{40},
  11202--11212\relax
\mciteBstWouldAddEndPuncttrue
\mciteSetBstMidEndSepPunct{\mcitedefaultmidpunct}
{\mcitedefaultendpunct}{\mcitedefaultseppunct}\relax
\EndOfBibitem
\bibitem[Phillips and Corces(2009)]{Phillips2009}
J.~E. Phillips and V.~G. Corces, \emph{Cell}, 2009, \textbf{137},
  1194--1211\relax
\mciteBstWouldAddEndPuncttrue
\mciteSetBstMidEndSepPunct{\mcitedefaultmidpunct}
{\mcitedefaultendpunct}{\mcitedefaultseppunct}\relax
\EndOfBibitem
\bibitem[Tang \emph{et~al.}(2015)Tang, Luo, Li, Zheng, Zhu, Szalaj, Trzaskoma,
  Magalska, Wlodarczyk, Ruszczycki, Michalski, Piecuch, Wang, Wang, Tian,
  Penrad-Mobayed, Sachs, Ruan, Wei, Liu, Wilczynski, Plewczynski, Li, and
  Ruan]{Tang2015}
Z.~Tang, O.~J. Luo, X.~Li, M.~Zheng, J.~J. Zhu, P.~Szalaj, P.~Trzaskoma,
  A.~Magalska, J.~Wlodarczyk, B.~Ruszczycki, P.~Michalski, E.~Piecuch, P.~Wang,
  D.~Wang, S.~Z. Tian, M.~Penrad-Mobayed, L.~M. Sachs, X.~Ruan, C.~L. Wei,
  E.~T. Liu, G.~M. Wilczynski, D.~Plewczynski, G.~Li and Y.~Ruan, \emph{Cell},
  2015, \textbf{163}, 1611--1627\relax
\mciteBstWouldAddEndPuncttrue
\mciteSetBstMidEndSepPunct{\mcitedefaultmidpunct}
{\mcitedefaultendpunct}{\mcitedefaultseppunct}\relax
\EndOfBibitem
\bibitem[Oti \emph{et~al.}(2016)Oti, Falck, Huynen, and Zhou]{Oti2016}
M.~Oti, J.~Falck, M.~A. Huynen and H.~Zhou, \emph{BMC Genomics}, 2016,
  \textbf{17}, 252\relax
\mciteBstWouldAddEndPuncttrue
\mciteSetBstMidEndSepPunct{\mcitedefaultmidpunct}
{\mcitedefaultendpunct}{\mcitedefaultseppunct}\relax
\EndOfBibitem
\bibitem[Brackley \emph{et~al.}(2017)Brackley, Johnson, Michieletto, Morozov,
  Nicodemi, Cook, and Marenduzzo]{Brackley2017prl}
C.~Brackley, J.~Johnson, D.~Michieletto, A.~Morozov, M.~Nicodemi, P.~Cook and
  D.~Marenduzzo, \emph{Phys. Rev. Lett.}, 2017, \textbf{119}, 138101\relax
\mciteBstWouldAddEndPuncttrue
\mciteSetBstMidEndSepPunct{\mcitedefaultmidpunct}
{\mcitedefaultendpunct}{\mcitedefaultseppunct}\relax
\EndOfBibitem
\bibitem[Davidson \emph{et~al.}(2016)Davidson, Goetz, Zaczek, Molodtsov, {Huis
  in 't Veld}, Weissmann, Litos, Cisneros, Ocampo-Hafalla, Ladurner, Uhlmann,
  Vaziri, and Peters]{Davidson2016}
I.~F. Davidson, D.~Goetz, M.~P. Zaczek, M.~I. Molodtsov, P.~J. {Huis in 't
  Veld}, F.~Weissmann, G.~Litos, D.~A. Cisneros, M.~Ocampo-Hafalla,
  R.~Ladurner, F.~Uhlmann, A.~Vaziri and J.~Peters, \emph{EMBO J.}, 2016,
  \textbf{35}, 2671--2685\relax
\mciteBstWouldAddEndPuncttrue
\mciteSetBstMidEndSepPunct{\mcitedefaultmidpunct}
{\mcitedefaultendpunct}{\mcitedefaultseppunct}\relax
\EndOfBibitem
\bibitem[Yamamoto and Schiessel(2017)]{Yamamoto2017}
T.~Yamamoto and H.~Schiessel, \emph{Phys. Rev. E}, 2017, \textbf{96},
  1--4\relax
\mciteBstWouldAddEndPuncttrue
\mciteSetBstMidEndSepPunct{\mcitedefaultmidpunct}
{\mcitedefaultendpunct}{\mcitedefaultseppunct}\relax
\EndOfBibitem
\bibitem[Higashi \emph{et~al.}(2021)Higashi, Pobegalov, Tang, Molodtsov, and
  Uhlmann]{Higashi2021}
T.~L. Higashi, G.~Pobegalov, M.~Tang, M.~I. Molodtsov and F.~Uhlmann,
  \emph{eLife}, 2021, \textbf{10}, 1--35\relax
\mciteBstWouldAddEndPuncttrue
\mciteSetBstMidEndSepPunct{\mcitedefaultmidpunct}
{\mcitedefaultendpunct}{\mcitedefaultseppunct}\relax
\EndOfBibitem
\bibitem[Ryu \emph{et~al.}(2021)Ryu, Bouchoux, Liu, Kim, Minamino, de~Groot,
  Katan, Bonato, Marenduzzo, Michieletto, Uhlmann, and Dekker]{Ryu2020}
J.-K. Ryu, C.~Bouchoux, H.~W. Liu, E.~Kim, M.~Minamino, R.~de~Groot, A.~J.
  Katan, A.~Bonato, D.~Marenduzzo, D.~Michieletto, F.~Uhlmann and C.~Dekker,
  \emph{Sci. Adv.}, 2021, \textbf{7}, 1--10\relax
\mciteBstWouldAddEndPuncttrue
\mciteSetBstMidEndSepPunct{\mcitedefaultmidpunct}
{\mcitedefaultendpunct}{\mcitedefaultseppunct}\relax
\EndOfBibitem
\bibitem[Marko \emph{et~al.}(2019)Marko, {De Los Rios}, Barducci, and
  Gruber]{Marko2019}
J.~F. Marko, P.~{De Los Rios}, A.~Barducci and S.~Gruber, \emph{Nucleic Acids
  Research}, 2019, \textbf{47}, 6956--6972\relax
\mciteBstWouldAddEndPuncttrue
\mciteSetBstMidEndSepPunct{\mcitedefaultmidpunct}
{\mcitedefaultendpunct}{\mcitedefaultseppunct}\relax
\EndOfBibitem
\bibitem[Diebold-Durand \emph{et~al.}(2017)Diebold-Durand, Lee, Ruiz~Avila,
  Noh, Shin, Im, Bock, B{\"u}rmann, Durand, Basfeld, and et~al.]{DieboldD2017}
M.~L. Diebold-Durand, H.~Lee, L.~B. Ruiz~Avila, H.~Noh, H.~C. Shin, H.~Im,
  F.~P. Bock, F.~B{\"u}rmann, A.~Durand, A.~Basfeld and et~al.,
  \emph{{Molecular Cell}}, 2017, \textbf{67}, 334--347\relax
\mciteBstWouldAddEndPuncttrue
\mciteSetBstMidEndSepPunct{\mcitedefaultmidpunct}
{\mcitedefaultendpunct}{\mcitedefaultseppunct}\relax
\EndOfBibitem
\bibitem[Terakawa \emph{et~al.}(2017)Terakawa, Bisht, Eeftens, Dekker, Haering,
  and Greene]{Terakawa2017}
T.~Terakawa, S.~Bisht, J.~M. Eeftens, C.~Dekker, C.~H. Haering and E.~C.
  Greene, \emph{Science}, 2017, \textbf{676}, eaan6516\relax
\mciteBstWouldAddEndPuncttrue
\mciteSetBstMidEndSepPunct{\mcitedefaultmidpunct}
{\mcitedefaultendpunct}{\mcitedefaultseppunct}\relax
\EndOfBibitem
\bibitem[Takaki \emph{et~al.}(2021)Takaki, Dey, Shi, and
  Thirumalai]{Takaki2020}
R.~Takaki, A.~Dey, G.~Shi and D.~Thirumalai, \emph{Nature Communications},
  2021, \textbf{12}, 5865\relax
\mciteBstWouldAddEndPuncttrue
\mciteSetBstMidEndSepPunct{\mcitedefaultmidpunct}
{\mcitedefaultendpunct}{\mcitedefaultseppunct}\relax
\EndOfBibitem
\bibitem[Nichols and G(2018)]{Nichols2018}
M.~H. Nichols and C.~V. G, \emph{{Nature Structural \& Molecular Biology}},
  2018, \textbf{25}, 906--901\relax
\mciteBstWouldAddEndPuncttrue
\mciteSetBstMidEndSepPunct{\mcitedefaultmidpunct}
{\mcitedefaultendpunct}{\mcitedefaultseppunct}\relax
\EndOfBibitem
\bibitem[Kschonsak \emph{et~al.}(2017)Kschonsak, Merkel, Bisht, Metz, Rybin,
  Hassler, and Haering]{Kschonsak2017}
M.~Kschonsak, F.~Merkel, S.~Bisht, J.~Metz, V.~Rybin, M.~Hassler and C.~H.
  Haering, \emph{Cell}, 2017, \textbf{171}, 588--600.e24\relax
\mciteBstWouldAddEndPuncttrue
\mciteSetBstMidEndSepPunct{\mcitedefaultmidpunct}
{\mcitedefaultendpunct}{\mcitedefaultseppunct}\relax
\EndOfBibitem
\bibitem[Nomidis \emph{et~al.}(2021)Nomidis, Carlon, Gruber, and
  F]{Nomidis2021}
S.~K. Nomidis, E.~Carlon, S.~Gruber and M.~J. F, \emph{bioRxiv}, 2021,
  2021.03.15.435506\relax
\mciteBstWouldAddEndPuncttrue
\mciteSetBstMidEndSepPunct{\mcitedefaultmidpunct}
{\mcitedefaultendpunct}{\mcitedefaultseppunct}\relax
\EndOfBibitem
\bibitem[Ryu \emph{et~al.}(2019)Ryu, Katan, van~der Sluis, Wisse, de~Groot,
  Haering, and Dekker]{RyuPre2019}
J.~K. Ryu, A.~J. Katan, E.~O. van~der Sluis, T.~Wisse, R.~de~Groot, C.~Haering
  and C.~Dekker, \emph{bioRxiv}, 2019,  2019.12.13.867358\relax
\mciteBstWouldAddEndPuncttrue
\mciteSetBstMidEndSepPunct{\mcitedefaultmidpunct}
{\mcitedefaultendpunct}{\mcitedefaultseppunct}\relax
\EndOfBibitem
\bibitem[Kamada \emph{et~al.}(2017)Kamada, Su\'etsugu, Takada, Miyata, and
  Hirano]{Kamada2017}
K.~Kamada, M.~Su\'etsugu, H.~Takada, M.~Miyata and T.~Hirano, \emph{Structure},
  2017, \textbf{25}, 603--616\relax
\mciteBstWouldAddEndPuncttrue
\mciteSetBstMidEndSepPunct{\mcitedefaultmidpunct}
{\mcitedefaultendpunct}{\mcitedefaultseppunct}\relax
\EndOfBibitem
\bibitem[Kim \emph{et~al.}(2020)Kim, Kerssemakers, Shaltiel, Haering, and
  Dekker]{Kim2020}
E.~Kim, J.~Kerssemakers, I.~A. Shaltiel, C.~H. Haering and C.~Dekker,
  \emph{Nature}, 2020, \textbf{579}, 438--442\relax
\mciteBstWouldAddEndPuncttrue
\mciteSetBstMidEndSepPunct{\mcitedefaultmidpunct}
{\mcitedefaultendpunct}{\mcitedefaultseppunct}\relax
\EndOfBibitem
\bibitem[Cheng \emph{et~al.}(2015)Cheng, Heeger, Chaleil, Matthews, Stewart,
  Wright, Lim, Bates, and Uhlmann]{Cheng2015}
T.~M. Cheng, S.~Heeger, R.~A. Chaleil, N.~Matthews, A.~Stewart, J.~Wright,
  C.~Lim, P.~A. Bates and F.~Uhlmann, \emph{eLife}, 2015, \textbf{4},
  1--22\relax
\mciteBstWouldAddEndPuncttrue
\mciteSetBstMidEndSepPunct{\mcitedefaultmidpunct}
{\mcitedefaultendpunct}{\mcitedefaultseppunct}\relax
\EndOfBibitem
\bibitem[Gibcus \emph{et~al.}(2018)Gibcus, Samejima, Goloborodko, Samejima,
  Naumova, Nuebler, Kanemaki, Xie, Paulson, Earnshaw, Mirny, and
  Dekker]{Gibcus2018}
J.~H. Gibcus, K.~Samejima, A.~Goloborodko, I.~Samejima, N.~Naumova, J.~Nuebler,
  M.~T. Kanemaki, L.~Xie, J.~R. Paulson, W.~C. Earnshaw, L.~A. Mirny and
  J.~Dekker, \emph{Science}, 2018, \textbf{359}, 6376\relax
\mciteBstWouldAddEndPuncttrue
\mciteSetBstMidEndSepPunct{\mcitedefaultmidpunct}
{\mcitedefaultendpunct}{\mcitedefaultseppunct}\relax
\EndOfBibitem
\bibitem[Gerguri \emph{et~al.}(2021)Gerguri, Fu, Kakui, Khatri, Barrington,
  Bates, and Uhlmann]{Gerguri2021}
T.~Gerguri, X.~Fu, Y.~Kakui, B.~S. Khatri, C.~Barrington, P.~A. Bates and
  F.~Uhlmann, \emph{Nucleic acids research}, 2021, \textbf{49},
  1294--1312\relax
\mciteBstWouldAddEndPuncttrue
\mciteSetBstMidEndSepPunct{\mcitedefaultmidpunct}
{\mcitedefaultendpunct}{\mcitedefaultseppunct}\relax
\EndOfBibitem
\bibitem[Pradhan \emph{et~al.}(2021)Pradhan, Barth, Kim, Davidson, Bauer, van
  Laar, Yang, Ryu, van~der Torre, Peters, and Dekker]{Pradhan2021}
B.~Pradhan, R.~Barth, E.~Kim, I.~F. Davidson, B.~Bauer, T.~van Laar, W.~Yang,
  J.-K. Ryu, J.~van~der Torre, J.-M. Peters and C.~Dekker, \emph{bioRxiv},
  2021\relax
\mciteBstWouldAddEndPuncttrue
\mciteSetBstMidEndSepPunct{\mcitedefaultmidpunct}
{\mcitedefaultendpunct}{\mcitedefaultseppunct}\relax
\EndOfBibitem
\bibitem[Ryu \emph{et~al.}(2020)Ryu, Rah, Janissen, Kerssemakers, and
  Dekker]{Ryu2021stepsize}
J.-K. Ryu, S.-H. Rah, R.~Janissen, J.~W.~J. Kerssemakers and C.~Dekker,
  \emph{bioRxiv}, 2020,  https://doi.org/10.1101/2020.11.04.368506\relax
\mciteBstWouldAddEndPuncttrue
\mciteSetBstMidEndSepPunct{\mcitedefaultmidpunct}
{\mcitedefaultendpunct}{\mcitedefaultseppunct}\relax
\EndOfBibitem
\bibitem[Nasmyth and Haering(2009)]{Nasmyth2009}
K.~Nasmyth and C.~H. Haering, \emph{Annu. Rev. Genet.}, 2009, \textbf{43},
  525--558\relax
\mciteBstWouldAddEndPuncttrue
\mciteSetBstMidEndSepPunct{\mcitedefaultmidpunct}
{\mcitedefaultendpunct}{\mcitedefaultseppunct}\relax
\EndOfBibitem
\bibitem[Murayama \emph{et~al.}(2018)Murayama, Samora, Kurokawa, Iwasaki, and
  Uhlmann]{Murayama2018}
Y.~Murayama, C.~P. Samora, Y.~Kurokawa, H.~Iwasaki and F.~Uhlmann, \emph{Cell},
  2018, \textbf{172}, 465--477.e15\relax
\mciteBstWouldAddEndPuncttrue
\mciteSetBstMidEndSepPunct{\mcitedefaultmidpunct}
{\mcitedefaultendpunct}{\mcitedefaultseppunct}\relax
\EndOfBibitem
\bibitem[Piskadlo \emph{et~al.}(2017)Piskadlo, Tavares, and
  Oliveira]{Piskadlo2017a}
E.~Piskadlo, A.~Tavares and R.~A. Oliveira, \emph{Elife}, 2017, \textbf{6},
  1--22\relax
\mciteBstWouldAddEndPuncttrue
\mciteSetBstMidEndSepPunct{\mcitedefaultmidpunct}
{\mcitedefaultendpunct}{\mcitedefaultseppunct}\relax
\EndOfBibitem
\bibitem[Brackley \emph{et~al.}(2017)Brackley, Liebchen, Michieletto, Mouvet,
  Cook, and Marenduzzo]{Brackley2017}
C.~A. Brackley, B.~Liebchen, D.~Michieletto, F.~Mouvet, P.~R. Cook and
  D.~Marenduzzo, \emph{Biophys J.}, 2017, \textbf{112}, 1085--1093\relax
\mciteBstWouldAddEndPuncttrue
\mciteSetBstMidEndSepPunct{\mcitedefaultmidpunct}
{\mcitedefaultendpunct}{\mcitedefaultseppunct}\relax
\EndOfBibitem
\bibitem[Bustamante \emph{et~al.}(1994)Bustamante, Marko, Siggia, and
  Smith]{Bustamante1994}
C.~Bustamante, J.~F. Marko, E.~Siggia and S.~B. Smith, \emph{Science}, 1994,
  \textbf{265}, 5--6\relax
\mciteBstWouldAddEndPuncttrue
\mciteSetBstMidEndSepPunct{\mcitedefaultmidpunct}
{\mcitedefaultendpunct}{\mcitedefaultseppunct}\relax
\EndOfBibitem
\bibitem[Plimpton(1995)]{Plimpton1995a}
S.~Plimpton, \emph{J. Comp. Phys.}, 1995, \textbf{117}, 1--19\relax
\mciteBstWouldAddEndPuncttrue
\mciteSetBstMidEndSepPunct{\mcitedefaultmidpunct}
{\mcitedefaultendpunct}{\mcitedefaultseppunct}\relax
\EndOfBibitem
\bibitem[Kimura and Hirano(1997)]{Kimura1997}
K.~Kimura and T.~Hirano, \emph{Cell}, 1997, \textbf{90}, 625--634\relax
\mciteBstWouldAddEndPuncttrue
\mciteSetBstMidEndSepPunct{\mcitedefaultmidpunct}
{\mcitedefaultendpunct}{\mcitedefaultseppunct}\relax
\EndOfBibitem
\bibitem[Kimura \emph{et~al.}(1999)Kimura, Rybenkov, Crisona, Hirano, and
  Cozzarelli]{Kimura1999}
K.~Kimura, V.~V. Rybenkov, N.~J. Crisona, T.~Hirano and N.~R. Cozzarelli,
  \emph{Cell}, 1999, \textbf{98}, 239--248\relax
\mciteBstWouldAddEndPuncttrue
\mciteSetBstMidEndSepPunct{\mcitedefaultmidpunct}
{\mcitedefaultendpunct}{\mcitedefaultseppunct}\relax
\EndOfBibitem
\bibitem[Kim \emph{et~al.}(2021)Kim, Gonzalez, Pradhan, {Van Der Torre}, and
  Dekker]{Kim2021}
E.~Kim, A.~M. Gonzalez, B.~Pradhan, J.~{Van Der Torre} and C.~Dekker,
  \emph{bioRxiv}, 2021,  2021.05.15.444164\relax
\mciteBstWouldAddEndPuncttrue
\mciteSetBstMidEndSepPunct{\mcitedefaultmidpunct}
{\mcitedefaultendpunct}{\mcitedefaultseppunct}\relax
\EndOfBibitem
\bibitem[Nomidis \emph{et~al.}(2019)Nomidis, Caraglio, Laleman, Phillips,
  Skoruppa, and Carlon]{Nomidis2019}
S.~K. Nomidis, M.~Caraglio, M.~Laleman, K.~Phillips, E.~Skoruppa and E.~Carlon,
  \emph{Phys. Rev. E}, 2019,  022402\relax
\mciteBstWouldAddEndPuncttrue
\mciteSetBstMidEndSepPunct{\mcitedefaultmidpunct}
{\mcitedefaultendpunct}{\mcitedefaultseppunct}\relax
\EndOfBibitem
\end{mcitethebibliography}
\bibliographystyle{rsc}

\end{document}